\documentclass[11pt]{article}
\usepackage{mathpazo}
\usepackage[latin9]{inputenc}
\usepackage[a4paper]{geometry}
\usepackage{units}
\usepackage{amsmath}
\usepackage{amssymb}

\makeatletter
\@ifundefined{date}{}{\date{}}
\usepackage{graphicx}
\usepackage{color}
\usepackage{calrsfs}
\usepackage[outerbars]{changebar}
\newcommand{\abs}[1]{\lvert#1\rvert}
\newcommand{\avg}[1]{\langle #1 \rangle}
\usepackage{units}

\makeatother

\begin{document}

\title{Universality in Uncertainty Relations\\
for a Quantum Particle}

\author{Spiros Kechrimparis\thanks{sk864@york.ac.uk}\phantom{x}and Stefan
Weigert\thanks{stefan.weigert@york.ac.uk} \\
 Department of Mathematics, University of York \\
 York, YO10 5DD, United Kingdom\\
\date{22 June 2016}}
\maketitle
\begin{abstract}
A general theory of preparational uncertainty relations for a quantum
particle in one spatial dimension is developed. We derive conditions
which determine whether a given smooth function of the particle's
variances and its covariance is bounded from below. Whenever a global
minimum exists, an uncertainty relation has been obtained. The squeezed
number states of a harmonic oscillator are found to be \emph{universal}:
no other pure or mixed states will saturate any such relation. Geometrically,
we identify a \emph{convex uncertainty region }in the space of second
moments which is bounded by the inequality derived by Robertson and
Schrödinger. Our approach provides a unified perspective on existing
uncertainty relations for a single continuous variable, and it leads
to new inequalities for second moments which can be checked experimentally.
\end{abstract}
\global\long\def\bk#1#2{\langle#1|#2\rangle}

\global\long\def\kb#1#2{|#1\rangle\,\langle#2|}

\global\long\def\braket#1#2{\langle#1|#2\rangle}

\global\long\def\ket#1{|#1\rangle}

\global\long\def\bra#1{\langle#1|}

\global\long\def\c#1{\mathbb{C}^{#1}}

\global\long\def\abs#1{\mid#1\mid}

\global\long\def\avg#1{\langle#1\rangle}

\section{Introduction and main result}

Inspired by Heisenberg's analysis \cite{heisenberg27} of Compton
scattering, Kennard \cite{kennard27} proved the preparational uncertainty
relation 
\begin{equation}
\Delta p\,\Delta q\geq\frac{\hbar}{2}\,,\label{eq: Heisenbergs inequality}
\end{equation}
for the standard deviations $\Delta p$ and $\Delta q$ of momentum
and position of a quantum particle with a single spatial degree of
freedom. Experimentally, they are determined by measurements performed
on an ensemble of systems prepared in a specific state $\ket{\psi}$.
The only states which saturate the bound (\ref{eq: Heisenbergs inequality})
are \emph{squeezed states} with a \emph{real} squeezing parameter
\cite{plebanski56,mollow+67,stoler70} (we follow the review \cite{dodonov02}
regarding the naming of squeezed states). Squeezed states are conceptually
important since they achieve the best possible localization of a quantum
particle in phase-space, and they are easily visualized by ``ellipses
of uncertainty''. Each squeezed state may be displaced rigidly in
phase space without affecting the value of the variances, resulting
in a three-parameter family of states saturating the lower bound \eqref{eq: Heisenbergs inequality}.

Not many other uncertainty relations are known. The \emph{sum} of
the position and momentum variances is bounded \cite{trifonov01,busch+13}
according to the relation 

\begin{equation}
\Delta^{2}p+\Delta^{2}q\geq\hbar\,,\label{eq: sum inequality}
\end{equation}
which holds in a system of units where the physical dimensions of
both position and momentum equal $\sqrt{\hbar}$. Only the ground
state of a harmonic oscillator with unit mass and frequency saturates
this inequality (ignoring rigid displacements in phase-space). The
Robertson-Schrödinger (RS) inequality \cite{robertson29,schroedinger30},
\begin{equation}
\Delta^{2}p\,\Delta^{2}q-C_{pq}^{2}\geq\frac{\hbar^{2}}{4}\,,\label{eq: RS inequality}
\end{equation}
sharpens Heisenberg's inequality \eqref{eq: Heisenbergs inequality}
by including the covariance $C_{pq}$ defined in Eq.~\eqref{eq: correlation term}.
Eq. \eqref{eq: RS inequality} is saturated by the two-parameter family
of\emph{ squeezed states} with a \emph{complex} squeezing parameter
\cite{dodonov02}, again ignoring phase-space displacements. The additional
free parameter describes the phase-space orientation of the uncertainty
ellipse which, in the previous case, was aligned with the position
and momentum axes. 

By introducing the observable $\hat{r}=-\hat{p}-\hat{q}$, which satisfies
the commutation relations $[\hat{q},\hat{r}]=[\hat{r},\hat{p}]=\hbar/i$,
one obtains a bound on the product of the variances of three pairwise
canonical observables, 
\begin{equation}
\Delta^{2}p\,\Delta^{2}q\,\Delta^{2}r\geq\left(\tau\frac{\hbar}{2}\right)^{3},\qquad\tau=\csc\left(\frac{2\pi}{3}\right)\equiv\sqrt{\frac{4}{3}}\,.\label{eq: triple product uncertainty}
\end{equation}
This \emph{triple product uncertainty} \emph{relation }has been found
only recently \cite{kechrimparis+14}. Ignoring phase-space translations,
only one state exists which achieves the minimum. Since the variance
of $\hat{r}$ is given by 
\begin{equation}
\Delta^{2}r=\Delta^{2}p+\Delta^{2}q+2C_{pq}\,,\label{eq: variance of r}
\end{equation}
the left-hand-side of (\ref{eq: triple product uncertainty}) can
also be considered as a function of the three second moments. 

The inequalities (\ref{eq: Heisenbergs inequality}) to (\ref{eq: triple product uncertainty})
and the search for their minima arise from one single mathematical
problem: 
\begin{quote}
\emph{Does a given smooth function of the second moments have a lower
bound}?\\
\emph{If so, which states will saturate the inequality}? 
\end{quote}
In this paper, we answer these questions for a quantum particle with
a single spatial degree of freedom by presenting a systematic approach
to studying uncertainty relations derived from smooth functions $f(\Delta^{2}p,\Delta^{2}q,C_{pq})$.
Proceeding in three steps we
\begin{enumerate}
\item identify a \emph{universal} set of states ${\cal E}$ which can possibly
minimize a given functional $f(\Delta^{2}p,\Delta^{2}q,C_{pq})$;
\item spell out conditions which determine the \emph{extrema} of the functional
$f$ as a subset of the universal set, ${\cal E}(f)\subseteq{\cal E}$;
if no admissible extrema exist, the functional has no lower bound; 
\item determine the set of states ${\cal M}(f)\subseteq{\cal E}(f)$ which
\emph{minimize} the functional $f$, leading to an uncertainty relation
in terms of the second moments. 
\end{enumerate}
The inequalities studied here will be \emph{preparational} in spirit:
they apply to scenarios in which the quantum state of the particle
$\ket{\psi}$ is fixed during the three separate runs of the measurements
required to determine the numerical values of the second moments.
These inequalities do not describe the limitations of measuring non-commuting
observables \emph{simultaneously}.

The paper is divided into two major sections. In Sec.~\ref{sec:Minimising-uncertainty}
we introduce uncertainty functionals and explain how to determine
their extrema and minima. The impatient reader may jump directly to
Sec.~\ref{sec:New-uncertainty-relations} where we derive new families
of uncertainty relations and determine the states minimizing them.
We conclude the paper with a summary and discuss further applications.

\section{Minimising uncertainty\label{sec:Minimising-uncertainty}}

To begin, we introduce the \emph{uncertainty functional }\cite{jackiw68}
\begin{equation}
J[\psi]=f\left(\Delta^{2}p,\Delta^{2}q,C_{pq}\right)-\lambda\left(\braket{\psi}{\psi}-1\right),\label{GUF}
\end{equation}
which sends each element $\ket{\psi}$ of the one-particle Hilbert
space ${\cal H}$ to a real number determined by the real differentiable
function $f(x_{1},x_{2},x_{3})$ of three variables. The Lagrange
multiplier $\lambda$ ensures the normalization of the states. The
variances of position and momentum are defined by 
\begin{equation}
\Delta^{2}p=\bra{\psi}\hat{p}^{2}\ket{\psi}-\bra{\psi}\hat{p}\ket{\psi}{}^{2}\,,\label{eq:define p variance}
\end{equation}
etc., and the covariance of position and momentum reads 
\begin{equation}
C_{pq}=\frac{1}{2}\bra{\psi}\left(\hat{p}\hat{q}+\hat{q}\hat{p}\right)\ket{\psi}-\bra{\psi}\hat{p}\ket{\psi}\bra{\psi}\hat{q}\ket{\psi}\,.\label{eq: correlation term}
\end{equation}
The second moments form the real, symmetric \emph{covariance matrix}
\begin{equation}
\mathbf{C}=\left(\begin{array}{cc}
\Delta^{2}p & C_{pq}\\
C_{pq} & \Delta^{2}q
\end{array}\right)\equiv\begin{pmatrix}x & w\\
w & y
\end{pmatrix}\,,
\end{equation}
with state-dependent matrix elements $x\equiv x(\psi)$ etc. The covariance
may take any finite real value, $w\in\mathbb{R}$, while the variances
of position and momentum take (finite)\emph{ positive} values only,
$x,y>0$. States of a quantum particle with \emph{vanishing} position
(or momentum) variance and \emph{diverging} momentum (or position)
variance are not taken into account since they only arise for non-normalizable
states which cannot be prepared experimentally. Nevertheless, position
(or momentum) eigenstates can be approximated arbitrarily well by
states within the set we consider.

It will be convenient to work with states in which the expectation
values of both momentum and position vanish, $\bra{\psi}\hat{q}\ket{\psi}=\bra{\psi}\hat{p}\ket{\psi}=0$.
This can be achieved by rigidly displacing the observables using the
unitary operator 
\begin{equation}
\hat{T}_{\alpha}=\exp\left[i\left(p_{0}\hat{q}-q_{0}\hat{p}\right)/\hbar\right]\,,\qquad\alpha=\frac{1}{\sqrt{2\hbar}}\left(q_{0}+ip_{0}\right)\,,\label{TranslOp}
\end{equation}
where $p_{0}=\bra{\psi}\hat{p}\ket{\psi}$, etc. This transformation
leaves invariant the values of the second moments \eqref{eq:define p variance}
and \eqref{eq: correlation term} and has thus no impact on the minimization
of the functional $J[\psi]$.

A lower bound of a functional $J[\psi]$ of the form (\ref{GUF})
will result in an uncertainty relation associated with the function
$f(x_{1},x_{2},x_{3})$. To determine such a bound, we apply a method
used in \cite{jackiw68,weigert96,kechrimparis+14} (see also \cite{busch+87,bialnicki+12,rudnicki12}).
First, we derive an eigenvalue equation for the extrema of the functional
$J[\psi]$ which also need to satisfy a set of consistency conditions
given in Sec.~\ref{subsec:Consistency-conditions}. Then, we introduce
a ``space of moments'' to visualize these results (Sec.~\ref{subsec:Geometry-of-extremal})
and, finally, we determine the minimizing states whenever the functional
is guaranteed to be bounded from below (see Sec.~\ref{subsec:Known-uncertainty-relations}
and Sec.~\ref{sec:New-uncertainty-relations}).

\subsection{Extrema of uncertainty functionals\label{subsec:Extrema-of-uncertainty}}

When comparing the values of the functional $J$ at the points $\ket{\psi}$
and $\ket{\psi+\varepsilon}\equiv\ket{\psi}+\varepsilon\ket e$, for
any unit vector $\ket e$ and a small parameter $\varepsilon$, we
find to first order that 
\begin{align}
J[\psi+\varepsilon] & -J[\psi]=\varepsilon D_{\varepsilon}J[\psi]+O\left(\varepsilon^{2}\right),
\end{align}
where the expression 
\begin{equation}
D_{\varepsilon}=\bra e\frac{\delta}{\delta\bra{\psi}}+\frac{\delta}{\delta\ket{\psi}}\ket e,
\end{equation}
denotes a Gâteaux derivative. If the functional $J[\psi]$ does not
change under this variation, 
\begin{align}
D_{\varepsilon}J[\psi]=\bra e\left(\frac{\delta}{\delta\bra{\psi}}f\left(x,y,w\right)-\lambda\ket{\psi}\right)+\text{c.c.}=0,
\end{align}
it has an extremum at the state $\ket{\psi}$. More explicitly, this
condition reads 
\begin{align}
\bra e\left(\frac{\partial f}{\partial x}\frac{\delta x}{\delta\bra{\psi}}+\frac{\partial f}{\partial y}\frac{\delta y}{\delta\bra{\psi}}+\frac{\partial f}{\partial w}\frac{\delta w}{\delta\bra{\psi}}-\lambda\ket{\psi}\right)+\text{c.c.}=0\,,
\end{align}
which should hold for arbitrary variations. Since the vector $\ket e$
and its dual $\bra e$ can be varied independently (just consider their position
representations $e^{*}(x)$ and $e(x)$), the expression in round
brackets must vanish identically which implies that the complex conjugate
term will also vanish. Using
\begin{equation}
\frac{\delta x}{\delta\bra{\psi}}\equiv\frac{\delta\Delta^{2}p}{\delta\bra{\psi}}=\frac{\delta\bra{\psi}\hat{p}^{2}\ket{\psi}}{\delta\bra{\psi}}=\hat{p}^{2}\ket{\psi}\,,
\end{equation}
a similar relation for $\delta y/\delta\bra{\psi}$, and the identity
\begin{equation}
\frac{\delta w}{\delta\bra{\psi}}\equiv\frac{1}{2}\left(\hat{q}\hat{p}+\hat{p}\hat{q}\right)\ket{\psi}\,,
\end{equation}
we arrive at an \emph{Euler-Lagrange}-type equation, 
\begin{equation}
\left(\frac{\partial f}{\partial x}\hat{p}^{2}+\frac{\partial f}{\partial y}\hat{q}^{2}+\frac{1}{2}\frac{\partial f}{\partial w}\left(\hat{q}\hat{p}+\hat{p}\hat{q}\right)-\lambda\right)\ket{\psi}=0.\label{eq: condition with lambda}
\end{equation}
The parameter $\lambda$ can be eliminated by multiplying this equation
with the bra $\bra{\psi}$ from the left and solving for $\lambda$;
substituting the value obtained back into Eq.~(\ref{eq: condition with lambda}),
one finds a nonlinear eigenvector-eigenvalue equation, 
\begin{equation}
\left(f_{x}\hat{p}^{2}+f_{y}\hat{q}^{2}+\frac{f_{w}}{2}\left(\hat{q}\hat{p}+\hat{p}\hat{q}\right)\right)\ket{\psi}=\left(f_{x}x+f_{y}y+f_{w}w\right)\ket{\psi}\,,\label{EigEq}
\end{equation}
using the standard shorthand for partial derivatives. 

Eq.~\eqref{EigEq} is our first result following from the approach
conceived in \cite{jackiw68}: the extrema of arbitrary smooth functions
of the second moments are encoded in an eigenvalue equation for a
Hermitean operator quadratic in position and momentum. However, the
equation is not linear in the state $\ket{\psi}$ because the quantities
$x,y,\ldots,f_{w}$ are functions of expectation values taken in the
yet unknown state. Previously, similar results had only been found
for specific uncertainty functionals such as the product of the standard
deviations or position and momentum.

Let us briefly illustrate the crucial features of Eq.~(\ref{EigEq})
in a simple case before systematically investigating its solutions.
For a function linear in $x,y,$ and $w$, the derivatives $f_{x},f_{y}$,
and $f_{w}$ will be fixed constant numbers. In this case, the operator
on the left-hand-side of (\ref{EigEq}) represents a quadratic form
in the position and momentum operators, falling into one of three
possible categories \cite{weigert02}. Up to a multiplicative constant,
the operator will be unitarily equivalent to the Hamiltonian of (i)
a harmonic oscillator with unit mass and frequency, $\hat{p}^{2}+\hat{q}^{2}$,
(ii) a free particle, $\hat{p}^{2}$, or (iii) an inverted harmonic
oscillator, $\hat{p}^{2}-\hat{q}^{2}$. In the first case, the spectrum
of the operator will be discrete and bounded from below (or above);
the spectra of the operators in the other two cases are continuous
which is tantamount to the absence of normalizable eigenstates. Thus,
a linear function $f(x,y,w)$ possesses a non-trivial bound only if
it gives rise to an operator in (\ref{EigEq}) which is unitarily
equivalent to the Hamiltonian of a harmonic oscillator. Generally,
our method will signal the absence of lower bounds corresponding to
the cases (ii) and (iii).

\subsection{Universality\label{subsec:Universality}}

To find a lower bound of the functional $J[\psi]$, we will determine
all its extrema and then pick those where $J[\psi]$ assumes its smallest
value. However, Eq.~(\ref{EigEq}) is not a standard eigenvalue equation:
even for a linear function $f$, the right-hand-side of (\ref{EigEq})
depends non-linearly on the as yet unknown state $\ket{\psi}$, and
if the function $f$ is non-linear, the operators on the left-hand-side
of the equation acquire state-dependent coefficients given by its
partial derivatives. 

Nevertheless, the eigenvalue problem can be solved systematically,
in a \emph{self-consistent} way. Initially, we treat the expectations
$x,y,w$, and $f_{x},f_{y},f_{w}$ in (\ref{EigEq}) as parameters
with given values, i.e. we ignore their dependence on the state $\ket{\psi}$.
The solutions $\ket{\psi(x,y,w)}$ will depend on these parameters
which means that the solutions must be checked for consistency since
the relations (\ref{eq:define p variance}) now require that $x=\bra{\psi(x,y,w)}\hat{p}^{2}\ket{\psi(x,y,w)}$,
etc. It may or may not be possible to satisfy these restrictions on
the parameters.

To begin, we write the operator on the left-hand side of (\ref{EigEq})
in matrix form,
\begin{equation}
(\hat{p}\,,\hat{q})\left(\begin{array}{cc}
f_{x} & f_{w}/2\\
f_{w}/2 & f_{y}
\end{array}\right)\left(\begin{array}{c}
\hat{p}\\
\hat{q}
\end{array}\right)\equiv\hat{\mathbf{z}}^{\top}\!\cdot\mathbf{F}\,\cdot\,\hat{\mathbf{z}}\,.\label{eq: define quadratic form via F}
\end{equation}
Williamson's theorem \cite{williamson36} ensures that any positive
or negative definite matrix can be mapped to a diagonal matrix by
conjugation with a symplectic matrix $\mathbf{\Sigma}$. We will assume
from now on that the matrix $\mathbf{F}$ is \emph{positive} definite.
The negative definite case is easily dealt with by considering $-f(x,y,w)$
instead of $f(x,y,w)$. Applied to the $2\times2$ matrix $\mathbf{F}$,
Williamson's result states that we can write 
\begin{equation}
\mathbf{\Sigma}^{\top}\!\cdot\mathbf{F}\cdot\mathbf{\Sigma}=\text{c\ensuremath{\mathbf{\,I}}\,},\qquad\mathbf{\Sigma}\in\text{Sp}(2,\mathbb{R})\,,\quad c>0\,,\label{eq: diagonalize F}
\end{equation}
where $\mathbf{I}$ is the identity matrix, whenever $\mathbf{F}>0$
holds. This requires both 
\begin{equation}
\det\,\mathbf{F}\equiv f_{x}f_{y}-f_{w}^{2}/4>0\quad\mbox{ and }\quad f_{x}>0\,,\label{eq: F > 0}
\end{equation}
implying that $f_{y}>0$ will hold, too. These requirements clearly
agree with the observations made for linear uncertainty functionals
$f(x,y,w)$: since the operators $\hat{p}^{2}$ and $\hat{p}^{2}-\hat{q}^{2}$
result in matrices $\mathbf{F}$ with zero or negative determinant,
the left-hand-side of (\ref{EigEq}) cannot be mapped to an oscillator
Hamiltonian by means of a symplectic transformation. 

A direct calculation shows that the matrix $\mathbf{F}$ is diagonalized
by the symplectic matrix $\Sigma=\left(\mathbf{S}_{\gamma}\mathbf{G}_{b}\right)^{-1}$,
where 

\begin{equation}
\textbf{G}_{b}=\left(\begin{array}{cc}
1 & 0\\
b & 1
\end{array}\right)\,,\mbox{\quad and}\quad\mathbf{S}_{\gamma}=\left(\begin{array}{cc}
e^{-\gamma} & 0\\
0 & e^{\gamma}
\end{array}\right),\label{eq: define G and S}
\end{equation}
with real parameters

\begin{equation}
b=\frac{f_{w}}{2f_{y}}\in\mathbb{R}\quad\mbox{and}\quad\gamma=\frac{1}{2}\ln\left(\frac{f_{y}}{\sqrt{\text{det}\,\textbf{F}}}\right)\,\in\mathbb{R},\label{eq: definitions of b and gamma}
\end{equation}
leading to $c=\sqrt{\det\,\mathbf{F}}$ in Eq.~\eqref{eq: diagonalize F}.
The symplectic matrices $\textbf{S}_{\gamma}$ and $\textbf{G}_{b}$
give rise to the Iwasawa (or $\mathcal{KAN}$) decomposition of the
matrix $\mathbf{\Sigma}^{-1}\in\text{Sp}(2,\mathbb{R})$ (cf. \cite{arvind+95},
for example) if they are written in opposite order and the parameter
$b$ is replaced by $be^{\gamma}$; the third factor happens to be
the identity. 

Next, we observe that the linear action of the matrices $\mathbf{G}$
and $\mathbf{S}$ on the canonical pair of operators $(\hat{p},\hat{q})^{\top}$
can be implemented by conjugation with suitable unitary operators,
known as \emph{metaplectic} operators \cite{arvind+95}. We have,
for example, 

\begin{equation}
\left(\begin{array}{cc}
1 & 0\\
b & 1
\end{array}\right)\,\left(\begin{array}{c}
\hat{p}\\
\hat{q}
\end{array}\right)=e^{ib\hat{p}^{2}/2\hbar}\left(\begin{array}{c}
\hat{p}\\
\hat{q}
\end{array}\right)e^{-ib\hat{p}^{2}/2\hbar}\,,\label{eq: action of G}
\end{equation}
or, in matrix notation,

\begin{equation}
\mathbf{G}_{b}\cdot\hat{\mathbf{z}}=\hat{G}_{b}\hat{\,\mathbf{z}}\,\hat{G}_{b}^{\dagger}\label{eq: Action of G - matrix vs}
\end{equation}
where the unitary operator

\begin{equation}
\hat{G}_{b}=e^{ib\hat{p}^{2}/2\hbar}\label{eq: gauge operator}
\end{equation}
describes a \emph{momentum gauge transformation}. Similarly, the squeeze
operator 
\begin{equation}
\hat{S}_{\gamma}=e^{i\gamma\left(\hat{q}\hat{p}+\hat{p}\hat{q}\right)/2\hbar}\,,\label{Squeeze}
\end{equation}
symplectically\emph{ scales} position and momentum according to 

\begin{equation}
\mathbf{S}_{\gamma}\cdot\hat{\mathbf{z}}=\hat{S}_{\gamma}\,\hat{\mathbf{z}}\,\hat{S}_{\gamma}^{\dagger}\,.\label{eq: action of S - matrix vs}
\end{equation}

With $\mathbf{\Sigma}^{-1}=\mathbf{S}_{\gamma}\mathbf{G}_{b}$ in
(\ref{eq: diagonalize F}), we rewrite (\ref{eq: define quadratic form via F})
as 

\begin{equation}
\hat{\mathbf{z}}^{\top}\!\cdot\mathbf{F}\,\cdot\,\hat{\mathbf{z}}=\sqrt{\det\mathbf{F}}\,(\mathbf{S}_{\gamma}\cdot\mathbf{G}_{b}\cdot\hat{\mathbf{z}})^{\top}\cdot(\mathbf{S}_{\gamma}\cdot\mathbf{G}_{b}\cdot\hat{\mathbf{z}})\,.
\end{equation}
Finally, using the identities (\ref{eq: Action of G - matrix vs})
and (\ref{eq: action of S - matrix vs}) and multiplying Eq.~(\ref{EigEq})
with the unitary $\hat{S}_{\gamma}^{\dagger}\hat{G}_{b}^{\dagger}$
from the left, the condition for the existence of extrema of the functional
$J[\psi]$ takes on the desired form, 
\begin{equation}
\frac{1}{2}\left(\hat{p}^{2}+\hat{q}^{2}\right)\ket{\psi(b,\gamma)}=\left(\frac{xf_{x}+yf_{y}+wf_{w}}{2\sqrt{\text{det}\,\mathbf{F}}}\right)\ket{\psi(b,\gamma)}\,.\label{HOeigen}
\end{equation}
Thus, the solutions
\begin{equation}
\ket{\psi(b,\gamma)}\equiv\hat{S}_{\gamma}^{\dagger}\hat{G}_{b}^{\dagger}\ket{\psi}\,
\end{equation}
 must be proportional to the eigenstates $\ket n,n\in\mathbb{N}_{0}$,
of a\emph{ unit oscillator}, i.e. a quantum mechanical oscillator
with unit mass and unit frequency. Equivalently, the candidates for
states extremizing the functional $J[\psi]$ are given by the family
of states, 
\begin{equation}
\ket{n(b,\gamma)}=\hat{G}_{b}\hat{S}_{\gamma}\ket n\,,\qquad b,\gamma\in\mathbb{R}\,,\quad n\in\mathbb{N}_{0}\,.\label{all Extremal States - initial}
\end{equation}
Upon rewriting the operator $\hat{S}_{\gamma}^{\dagger}\hat{G}_{b}^{\dagger}$
these states are seen to coincide with the\emph{ squeezed number states}
\cite{dodonov02}. As shown in in Appendix \ref{sec: An-operator-identity},
the product of a squeeze transformation $\hat{S}_{\gamma}$ (with
real parameter $\gamma$) and a momentum gauge transformation $\hat{G}_{b}$
equals 

\begin{equation}
\hat{G}_{b}\hat{S}_{\gamma}=\hat{S}(\xi)\hat{R}(\chi)\,,\label{eq: BCH identity: GS =00003D SR}
\end{equation}
i.e. the product of a rotation in phase space,

\begin{equation}
\hat{R}(\chi)=e^{i\chi\hat{a}^{\dagger}\hat{a}}\,\qquad\chi\in[0,2\pi)\,,
\end{equation}
and a squeeze transformation (with complex $\xi$) along a line with
inclination $\theta$, 

\begin{equation}
\hat{S}(\xi)=e^{\frac{1}{2}\left(\xi\hat{a}^{\dagger2}-\overline{\xi}\hat{a}^{2}\right)}\,,\qquad\xi=re^{i\theta}\in\mathbb{C}\,.
\end{equation}

Summarizing our findings, we draw two conclusions:
\begin{enumerate}
\item The complete set of solutions of Eq.~(\ref{HOeigen}) coincides with
the\emph{ squeezed number states},
\begin{equation}
{\cal E}=\bigcup_{n=0}^{\infty}{\cal E}_{n}\equiv\bigcup_{n=0}^{\infty}\left\{ \ket{n(\alpha,\xi)}=\hat{T}_{\alpha}\hat{S}(\xi)\ket n\,,\alpha,\xi\in\mathbb{C}\right\} \,,\label{eq: all extremal states - final}
\end{equation}
where non-zero expectation values of position and momentum have been
reintroduced via the translation operator $\hat{T}_{\alpha}$ (see
Eq.~(\ref{TranslOp})) and irrelevant constant phases have been suppressed.
\item The value of the right-hand-side of Eq.~(\ref{HOeigen}) can take
only specific values, 
\begin{equation}
\frac{xf_{x}+yf_{y}+wf_{w}}{2\sqrt{\text{det}\,\mathbf{F}}}=\left(n+\frac{1}{2}\right)\hbar\,,\qquad n\in\mathbb{N}_{0}\,,\label{EigCond}
\end{equation}
given by the eigenvalues of the unit oscillator. This relation constrains
the state-dependent quantities of the left-hand-side which needs to
be checked for consistency, just as Eq.~(\ref{eq: F > 0}) does.
\end{enumerate}
We have thus obtained our second main result. The extrema ${\cal E}$
of an arbitrary functional $J[\psi]$ characterized by a function
$f(x,y,w)$ are necessarily \emph{squeezed number states}, a set which
is \emph{independent} of the function at hand. In other words, the
set ${\cal E}$ containing all the states which may arise as minima
of an uncertainty functional $J[\psi]$, is \emph{universal}. The
minima of any functional must be a subset ${\cal E}(f)\subseteq{\cal E}$
which will depend explicitly on the function $f(x,y,w)$, determined
by the consistency conditions to be studied next.

\subsection{Consistency conditions\label{subsec:Consistency-conditions} }

We now spell out the conditions which must be satisfied by the states
$\ket{n(b,\gamma)}$ in (\ref{all Extremal States - initial}) \textendash{}
or, equivalently, the states $\ket{n(\alpha,\xi)}$ in (\ref{eq: all extremal states - final})
\textendash{} to qualify as extrema for a specific functional $J[\psi]$:
\begin{enumerate}
\item Recalling that $x\equiv\Delta^{\!2}p$, etc., the relations 
\begin{equation}
x=\bra{n(b,\gamma)}\hat{p}^{2}\ket{n(b,\gamma)}\,,\qquad y=\bra{n(b,\gamma)}\hat{q}^{2}\ket{n(b,\gamma)}\,,\label{eq: consistency for xy}
\end{equation}
and 
\begin{equation}
w=\frac{1}{2}\bra{n(b,\gamma)}\left(\hat{p}\hat{q}+\hat{q}\hat{p}\right)\ket{n(b,\gamma)}\,,\label{eq: consistency for w}
\end{equation}
represent three, generally nonlinear \emph{consistency equations }between
the second moments since the parameters $b$ and $\gamma$ are functions
of $x,y$ and $w$ (cf. Eq.~(\ref{eq: definitions of b and gamma})).
\item The values of the moments $x,y$ and $w$ must satisfy Eq.~\eqref{EigCond}.
\item The matrix $\mathbf{F}$ of the first derivatives must be positive
definite.
\end{enumerate}
Using \eqref{all Extremal States - initial}, \eqref{eq: Action of G - matrix vs}
and \eqref{eq: action of S - matrix vs}, the first consistency condition
in (\ref{eq: consistency for xy}) leads to
\begin{align}
x & =\bra{n(b,\gamma)}\hat{p}^{2}\ket{n(b,\gamma)}=e^{2\gamma}\bra n\hat{p}^{2}\ket n=e^{2\gamma}\left(n+\frac{1}{2}\right)\hbar\,,\quad n\in\mathbb{N}_{0}\,,
\end{align}
or, recalling the definition of $\gamma$ in (\ref{eq: definitions of b and gamma}),
\begin{equation}
x\sqrt{\text{det}\,\textbf{F}}=\left(n+\frac{1}{2}\right)\hbar f_{y}\,,\qquad n\in\mathbb{N}_{0}\,.\label{eq: x consistency}
\end{equation}
Similar calculations result in

\begin{equation}
y\sqrt{\text{det}\,\textbf{F}}=\left(n+\frac{1}{2}\right)\hbar f_{x}\,,\qquad n\in\mathbb{N}_{0}\,,\label{eq: y  consistency}
\end{equation}
and 
\begin{equation}
-2w\sqrt{\text{det}\,\textbf{F}}=\left(n+\frac{1}{2}\right)\hbar f_{w}\,,\qquad n\in\mathbb{N}_{0}\,,\label{eq: w consistency}
\end{equation}
respectively. These conditions may be expressed in matrix form, 

\begin{equation}
\frac{\text{{\bf F}}\cdot\text{{\bf C}}}{\sqrt{\text{det}\,\textbf{F}}}=\left(n+\frac{1}{2}\right)\hbar\,\mathbf{I},\qquad n\in\mathbb{N}_{0}\,,\label{eq: consistency matrix form}
\end{equation}
involving both the covariance matrix $\mathbf{C}$ and $\mathbf{F}$.

Taking the trace of the last relation shows that Eq.~\eqref{EigCond}
is satisfied automatically. With\-out specifying a function $f(x,y,w)$,
no conclusions can be drawn about the validity of Eqs.~(\ref{eq: x consistency}-\ref{eq: w consistency})
or the positive definiteness of the matrix $\mathbf{F}$.

\subsection{Geometry of extremal states\label{subsec:Geometry-of-extremal}}

\begin{figure}
\centering{}\def\svgwidth{0.7\textwidth}
\begin{center}
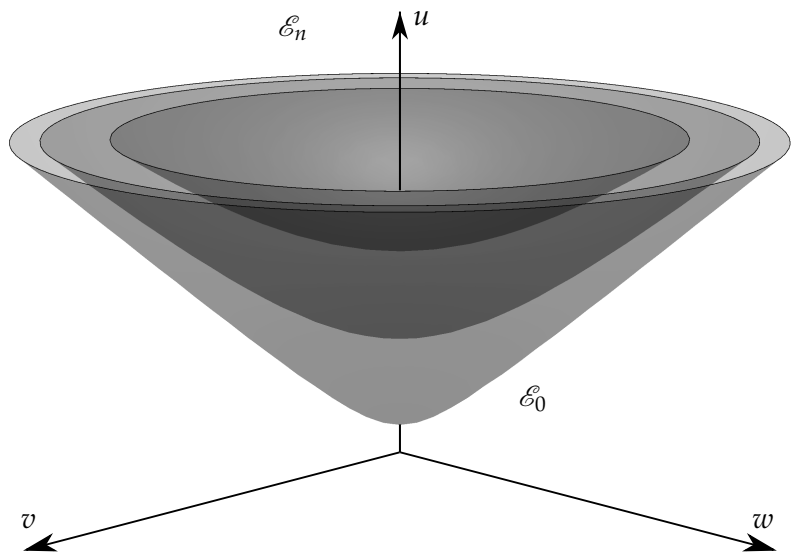
\end{center}\caption{\label{Fig. hyperboloidsinspaceofmoments} Space of (second) moments,
with points $(u,v,w)$: the extremal states of smooth functionals
$J[\psi]$ are located on a discrete set of nested hyperboloids ${\cal E}=\bigcup_{n=0}^{\infty}{\cal E}_{n}$
the first three of which are shown, using light ($n=0$), medium ($n=1$)
and dark shading ($n=2)$, respectively. The accessible \emph{uncertainty
region} for a quantum particle is given by the points on and inside
of the convex surface ${\cal E}_{0}:u^{2}-v^{2}-w^{2}=\hbar^{2}/4$
which coincides with the minima ${\cal M}(f^{RS})$ of the RS inequality,
i.e. squeezed states with minimal uncertainty. }
\end{figure}

We visualize the interplay of the consistency conditions by expressing
them in the form

\begin{equation}
xf_{x}=yf_{y}\,,\qquad xf_{w}=-2wf_{y}\,,\label{eq: alternative consistency I}
\end{equation}
and
\begin{equation}
xy-w^{2}=\left(n+\frac{1}{2}\right)^{2}\hbar^{2}\,,\qquad n\in\mathbb{N}_{0}\,,\label{eq: alternative consistency RS part}
\end{equation}
following easily from either (\ref{eq: x consistency}-\ref{eq: w consistency})
or (\ref{eq: consistency matrix form}). The third constraint is \emph{universal
}since it does not depend on the function $f(x,y,w)$. Using the variables
\[
u=\frac{1}{2}\left(x+y\right)>0\,,\quad v=\frac{1}{2}\left(x-y\right)\in\mathbb{R}\,,
\]
we define the three-dimensional \emph{space of }(\emph{second}) \emph{moments,}
with coordinates $(u,v,w)$. For each non-negative integer, the third
condition 
\begin{equation}
u^{2}-v^{2}-w^{2}=e_{n}^{2}\,,\qquad e_{n}=\left(n+\frac{1}{2}\right)\hbar\,,\quad n\in\mathbb{N}_{0}\,,\label{eq: uvw hyperboloid}
\end{equation}
determines one sheet of a two-sheeted hyperboloid, located in the
``upper'' half of the space of moments, i.e. $u>0$ and $v,w\in\mathbb{R}$
(cf. Fig.~\ref{Fig. hyperboloidsinspaceofmoments}). The points on
the $n$-th sheet, which intersects the $u$-axis at $u=+e_{n}$,
are in one-to-one correspondence with the squeezed states originating
from the number state $\ket n$, forming the set ${\cal E}_{n}$ in
\eqref{eq: all extremal states - final}. 

The consistency conditions \eqref{eq: alternative consistency I}
clearly depend on the function $f(x,y,w)$ at hand. The constraints
will only be satisfied for specific subsets ${\cal E}_{n}(f)$ of
points on the hyperboloids ${\cal E}_{n}$, resulting in the $f$-dependent
set of states 
\[
{\cal E}(f)=\bigcup_{n=0}^{\infty}{\cal E}_{n}(f)
\]
which contains all the candidates possibly minimizing the functional
$J[\psi]$. The candidate sets ${\cal E}(f)$ may depend on one or
two parameters, or contain isolated points only. If the consistency
conditions cannot be satisfied, then the functional $J[\psi]$ has
no lower bound. Furthermore, if the matrix $\mathbf{F}$ is not positive
definite for any of the states in ${\cal E}(f)$, the method makes
no predictions about the minima of the functional $J[\psi]$. We have,
however, not found any non-trivial cases of this behaviour.

Finally, we need to evaluate the functional $J[\psi]$ for all candidate
states ${\cal E}(f)$ and pick the smallest possible value. The states
achieving this minimum value constitute the solutions ${\cal M}(f)\subseteq{\cal E}(f)$
of the minimization problem. In their entirety, the minima ${\cal M}(f)$
may consist of isolated states or of sets depending on one or two
parameters. Usually, the states saturating the bound are located on
the sheet ${\cal E}_{0}$.

Let us briefly introduce the concept of a \emph{space of moments},
defined by the triples of numbers $(u,v,w)^{\top}\in\mathbb{R}^{3}$.
For $n=0$, Eq.~\eqref{eq: uvw hyperboloid} is equivalent to \eqref{eq: RS inequality}
which implies that not all points of this set can arise as moment
triples. The accessible part of the space, bounded by the extremal
hyperboloid ${\cal E}_{0}$ defined in Eq.~\eqref{eq: uvw hyperboloid},
is called the \emph{uncertainty region} (cf. Fig.~\ref{Fig. hyperboloidsinspaceofmoments}).
The boundary of an analogously defined uncertainty region for a quantum
spin $s$ \cite{dammeier+15} is not convex. The relation between
moment triples and the underlying pure or mixed states of quantum
particles is discussed elsewhere \cite{kechrimparis+16}. 

\subsection{Known uncertainty relations\label{subsec:Known-uncertainty-relations}}

To illustrate our approach we re-derive three of the four bounds mentioned
in the introduction: the\emph{ }uncertainty relations by \emph{Robertson-Schrödinger},
by \emph{Heisenberg-Kennard}, and the \emph{triple-product} inequality.

\emph{Robertson-Schrödinger uncertainty relation}. Defining
\begin{equation}
f^{RS}(x,y,w)=xy-w^{2}\,,\label{eq: RS function}
\end{equation}
the matrix of first-order derivatives associated with the quadratic
form (\ref{eq: define quadratic form via F}) is given by

\begin{equation}
\mathbf{F}=\begin{pmatrix}y & -w\\
-w & x
\end{pmatrix}\,,\label{eq: F for RS}
\end{equation}
and, interestingly, its determinant
\begin{equation}
\mbox{det}\,\mathbf{F}=xy-w^{2}\equiv f^{RS}(x,y,w)\label{eq: RS determinant}
\end{equation}
coincides with the original functional. At this point of the derivation,
it is not yet known whether the matrix $\mathbf{F}$ is strictly positive. 

The relations Eq.~(\ref{eq: alternative consistency I}) do not constraint
the parameters $x,y$, and $w$, since they are satisfied automatically,
leaving Eq.~(\ref{eq: alternative consistency RS part}) as the only
restriction. Since the left-hand-side of (\ref{eq: alternative consistency RS part})
coincides with the function $f^{RS}(x,y,w)$, all squeezed states
are candidates to minimize the RS functional, 
\begin{equation}
{\cal E}(f^{RS})={\cal E}\,.\label{eq: fRS candidates}
\end{equation}
Therefore, the function $f^{RS}$ comes with the largest possible
set of candidates to minimize it, given by the union of the sets ${\cal E}_{n}$
in Fig.~\ref{Fig. hyperboloidsinspaceofmoments}. The lower bound
on $f^{RS}$ now follows directly from combining (\ref{eq: alternative consistency RS part})
with \eqref{eq: RS function}, 
\begin{equation}
f^{RS}(x,y,w)=\left(n+\frac{1}{2}\right)^{2}\hbar^{2}\geq\frac{\hbar^{2}}{4}\,,\label{RSExtremal}
\end{equation}
reproducing the RS inequality. Eqs. \eqref{eq: RS determinant} and
\eqref{RSExtremal} imply that the determinant of $\mathbf{F}$ is
positive everywhere in the uncertainty region. 

The hyperboloid closest to the origin of the $(u,v,w)$-space provides
the states minimizing the function $f^{RS}$, 
\begin{equation}
{\cal M}(f^{RS})={\cal E}_{0}\,,\label{eq: fRS minima}
\end{equation}
i.e. the set of squeezed states based on the ground state $\ket 0$
of a unit oscillator. This is, of course, a two-parameter family since
the relations \eqref{eq: definitions of b and gamma} take the form
\begin{equation}
b=-\frac{z}{x}\,,\quad\mbox{and}\quad\gamma=\frac{1}{2}\ln\left(2x\right)\,,
\end{equation}
meaning that, with $x>0$ and $z\in\mathbb{R}$, both $b$ and $\gamma$
take indeed arbitrary real values. Thus, each squeezed state can be
reached and, when adding phase-space translations, we obtain the four-parameter
family of \emph{all} squeezed states as minima of $f^{RS}$: 
\begin{equation}
{\cal M}_{\alpha}(f^{RS})=\hat{T}_{\alpha}{\cal M}(f^{RS})\,.\label{eq: dislaced fRS minima}
\end{equation}
This property singles out the RS functional among all uncertainty
functionals. 

If an uncertainty functional associated to a function $f$, is different
from the RS functional, the first two consistency relations will,
in general, \emph{not} be satisfied automatically but impose non-trivial
constraints on the second moments. Therefore, the extrema of the functional
must be a proper subset of those of the RS functional, i.e. ${\cal E}(f)\subset{\cal E}$,
as the following example shows.

\begin{figure}
\def\svgwidth{0.7\textwidth}
\begin{center}
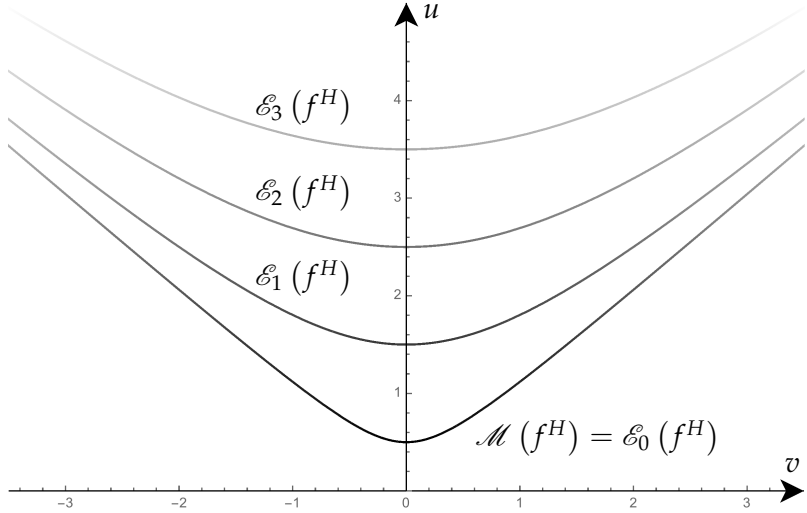
\end{center}

\caption{\label{fig:HB extrema}Hyperbolas in the $(u,v)$-plane through the
points $(e_{n},0)$, $n\in\mathbb{N}_{0}$, stemming from intersections
of the hyperboloids \eqref{eq: uvw hyperboloid} and the surfaces
defined by \eqref{eq: HB extrema}. The union of the hyperbolas defines
the set ${\cal E}(f^{H})$ which represents the location of all possible
minima of the function $f^{H}(x,y)$; the points on the ``lowest''
(darkest) hyperbola ${\cal {\cal E}}_{0}(f^{H})$ correspond to the
set of states ${\cal M}(f^{H})$ which saturate Heisenberg's uncertainty
relation.}
\end{figure}

\emph{Heisenberg's uncertainty relation}. Let us determine the minimum
of the product of the standard deviations $\Delta p$ and $\Delta q$
by considering the function 
\begin{equation}
f^{H}(x,y,w)=\sqrt{xy}\,.\label{eq: HB functional}
\end{equation}
Its partial derivatives satisfy 
\begin{equation}
2f_{x}^{H}=\sqrt{\frac{y}{x}}\,,\qquad2f_{y}^{H}=\sqrt{\frac{x}{y}}\,,\qquad f_{w}^{H}=0\,,\label{eq: HB consistency}
\end{equation}
resulting in a positive definite diagonal matrix $\mathbf{F}$, namely,
\begin{equation}
\mathbf{F}=\frac{1}{2}\begin{pmatrix}\sqrt{y/x} & 0\\
0 & \sqrt{x/y}
\end{pmatrix}\,,\quad\det\mathbf{F}=\frac{1}{4}>0\,.\label{eq: F for Heisenberg}
\end{equation}
Eqs.~(\ref{eq: x consistency}) and (\ref{eq: y  consistency}) collapse
into the conditions 
\begin{equation}
\sqrt{xy}=\left(n+\frac{1}{2}\right)\hbar\equiv e_{n}\,,\quad n\in\mathbb{N}_{0}\,,\label{eq: HB extrema}
\end{equation}
which determine the value of the product of the standard deviations
at the extrema of $f^{H}(x,y,w)$, labeled by the positive integers.
In the $(u,v,w)$-space, the intersections of the surfaces defined
by \eqref{eq: HB extrema} and the hyperboloids \eqref{eq: uvw hyperboloid}
consist of hyperbolas in the $(u,v)$-plane containing the points
$(e_{n},0)$, $n\in\mathbb{N}_{0}$. The union of these hyperbolas
define the set ${\cal E}(f^{H})$, corresponding to the potential
minima of the function $f^{H}(x,y)$ (cf. Fig.~\ref{fig:HB extrema}).
The third condition, Eq.~(\ref{eq: w consistency}), implies that
$w=0$. Combining \eqref{eq: HB extrema} with \eqref{eq: HB functional},
we obtain the bound 
\begin{equation}
f^{H}(x,y,w)=\left(n+\frac{1}{2}\right)\hbar\geq\frac{\hbar}{2}\,,\label{eq: HB bound}
\end{equation}
reproducing Heisenberg's uncertainty relation (\ref{eq: Heisenbergs inequality}). 

The family of states minimizing Heisenberg's uncertainty relation
is found by using the identity (\ref{eq: HB extrema}) for $n=0$
in the definition of the parameter $\gamma$ in (\ref{eq: definitions of b and gamma}),
leading to $f_{y}=x/\hbar$. Since the consistency conditions do not
impose any other condition on the variance $x$, it may take any positive
value implying that $\gamma\in\mathbb{R}$. Since $\mbox{f}_{w}=0$
leads to $b=0$, the set of states minimizing the Heisenberg's uncertainty
relation is given by squeezed states with real squeezing parameter, 

\begin{equation}
{\cal M}_{\alpha}(f^{H})=\hat{T}_{\alpha}{\cal E}(f^{H})\equiv\left\{ \hat{T}_{\alpha}\hat{S}_{\gamma}\ket 0\,,\,\alpha\in\mathbb{C},\gamma\in\mathbb{R}\right\} \,,
\end{equation}
where we have re-introduced arbitrary phase-space displacements.

\emph{Triple product inequality.} Using Eq.~(\ref{eq: variance of r}),
we see that we need to find the minimum of the expression

\begin{equation}
f^{T}(x,y,w)=xy\left(x+y+2w\right)
\end{equation}
in order to reproduce the triple product uncertainty relation (\ref{eq: triple product uncertainty}).
The first consistency condition in Eq.~(\ref{eq: alternative consistency I})
implies that $x=y$; using this identity in the second condition,
one finds 
\begin{equation}
x(w+x)(w+\frac{x}{2})=0\,.\label{eq: second condition Triple consistency}
\end{equation}
Recalling that variances are always non-zero for normalizable states,
$x>0$, the correlation $w$ must equal either $-x$ or $-x/2$. According
to (\ref{eq: variance of r}), the first case would imply $\Delta^{2}r\equiv0$,
which is impossible since the operator $\hat{r}$ has no normalizable
eigenstates. Therefore, using the solution $w=-x/2$ of (\ref{eq: second condition Triple consistency})
and $x=y$ in the third consistency condition, one finds that 
\begin{equation}
x^{2}=\frac{4}{3}\left(n+\frac{1}{2}\right)^{2}\hbar^{2}\,,\quad n\in\mathbb{N}_{0}\,,
\end{equation}
must hold. It is now straightforward to evaluate $f^{T}(x,y,w)$ at
its extrema to find its global minimum,
\[
f^{T}(x,y,w)=x^{3}=\left(\sqrt{\frac{4}{3}}\left(n+\frac{1}{2}\right)\hbar\right)^{3}\geq\left(\tau\frac{\hbar}{2}\right)^{3}\,,
\]
which reproduces (\ref{eq: triple product uncertainty}). It is easy
to confirm that the matrix $\mathbf{F}$ is positive definite with
determinant $\det\mathbf{F}=\hbar^{4}/3$. Since the minimum occurs
for $n=0$, the values of the second moments are given by
\begin{equation}
x=y=-2w=\frac{\hbar}{\sqrt{3}}\equiv\tau\frac{\hbar}{2}\,.\label{eq: values of 2nd moments tripleUR}
\end{equation}
These relations fix the values of the parameters in (\ref{eq: definitions of b and gamma}),

\begin{equation}
b=\frac{1}{2}\quad\mbox{and}\quad\gamma=\frac{1}{4}\ln\tau\,.\label{eq: tvalues of  b and gamma for TUR}
\end{equation}
Using (\ref{all Extremal States - initial}) or (\ref{eq: all extremal states - final})
we obtain one single state which saturates the triple uncertainty,
namely 
\begin{equation}
\ket{\Xi_{0}}\equiv\hat{G}_{\frac{1}{2}}\hat{S}_{\frac{1}{2}\ln\tau}\ket 0=\hat{S}_{\frac{i}{2}\ln3}\ket 0\,.\label{eq: Xi_0}
\end{equation}
If one includes rigid phase-space translations, the set of states
minimizing the triple uncertainty is finally given by the two-parameter
family 
\begin{equation}
{\cal M}_{\alpha}(f^{T})=\left\{ \hat{T}_{\alpha}\ket{\Xi_{0}}\,,\,\alpha\in\mathbb{C}\right\} \,,\label{eq: f^T minima}
\end{equation}
in agreement with \cite{kechrimparis+14}.

\begin{figure}
\def\svgwidth{0.7\textwidth}
\begin{center}
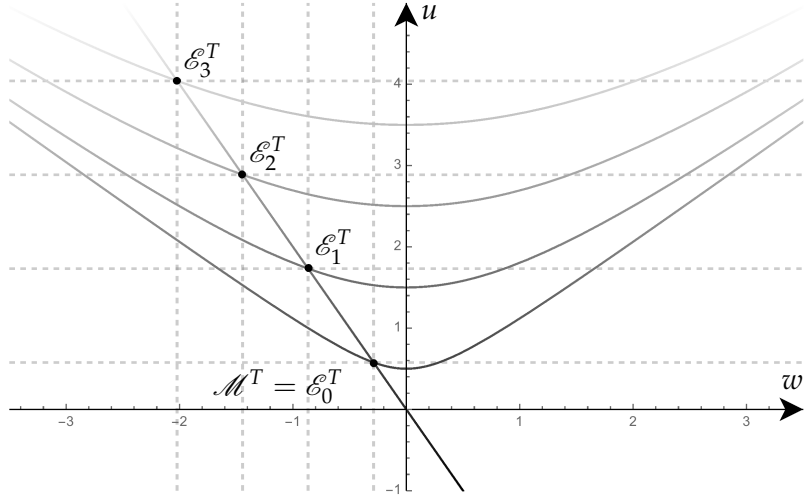
\end{center}

\caption{\label{fig: triple-uncertainty-visual} Candidate states ${\cal E}(f^{T})\equiv{\cal E}^{T}$
possibly minimizing the product of three variances $f^{T}(u,v,w)$,
represented by dots located on the intersections of the hyperboloids
\eqref{eq: uvw hyperboloid} and the planes defined by the consistency
conditions \eqref{eq: f^T candidates}; the point closest to the origin,
${\cal M}(f^{T})\equiv{\cal M}^{T}$, represents the state $\protect\ket{\Xi_{0}}$
achieving the minimum of the triple product uncertainty relation \eqref{eq: triple product uncertainty}
(and of any other $S_{3}$-invariant inequality associated with a
functional $f_{N}^{(3)}$ in \eqref{eq: def f(3)N}).}
\end{figure}

Geometrically, the state $\ket{\Xi_{0}}$ arises from the intersection
of the sequence of hyperboloids with the surfaces defined by 
\begin{equation}
u=\tau e_{n}\,,\quad v=0\,,\quad w=-\frac{1}{2}\tau e_{n}\,,\quad n\in\mathbb{N}_{0}\,.\label{eq: f^T candidates}
\end{equation}
The planes defined by constant values of $u$ have concentric circles
in common with the hyperboloids, and the vertical $uw$-plane (given
by $v=0$) intersects with each of the circles in two points only.
Finally, the condition on the variable $w$ selects a single one of
the points with the same value of $u$. According to \eqref{eq: f^T candidates},
the candidate states in $(u,v,w)$-space are located on a straight
line, 
\begin{equation}
{\cal E}(f^{T})=\left\{ \tau e_{n}\begin{pmatrix}1\\
0\\
-1/2
\end{pmatrix},n\in\mathbb{N}_{0}\right\} \,,\label{eq: triple candidate set}
\end{equation}
and the state $\ket{\Xi_{0}}$ corresponds to the point closest to
the origin (see Fig. (\ref{fig: triple-uncertainty-visual})).

\section{New uncertainty relations\label{sec:New-uncertainty-relations}}

\subsection{Generalizing known relations}

The \emph{linear combination }of second moments 
\begin{equation}
f^{L}(x,y,w)=\mu x+\nu y+2\lambda w\,,\qquad\mu,\nu,\lambda,\in\mathbb{R}\,,\label{eq: most general linear f}
\end{equation}
leads to the uncertainty relation 
\begin{equation}
\mu\Delta^{\!2}p+\nu\Delta^{\!2}q+2\lambda C_{pq}\geq\hbar\sqrt{\mu\nu-\lambda^{2}},\qquad\mu,\nu>0\,,\quad\mu\nu>\lambda^{2}\,.\label{eq:bound on most general f}
\end{equation}
The constraints on the parameters follow from the matrix $\mathbf{F}$
in \eqref{eq: define quadratic form via F} being strictly positive
definite. The consistency conditions (\ref{eq: alternative consistency I})
associated with $f^{L}$ relate both $y$ and $w$ to $x$ according
to 
\begin{equation}
y=\frac{\mu}{\nu}x\,,\qquad w=-\frac{\lambda}{\nu}x\,.
\end{equation}
 Then, Eq.~(\ref{eq: alternative consistency RS part}) simplifies
to 
\begin{equation}
\frac{\left(\mu\nu-\lambda^{2}\right)}{\nu^{2}}x^{2}=\left(n+\frac{1}{2}\right)^{2}\hbar^{2}=e_{n}^{2}\hbar^{2}\,,
\end{equation}
which is consistent due to $\det\mathbf{F}=\mu\nu-\lambda^{2}>0$.
Expressing the functional $f^{L}(x,y,w)$ in terms of $x$ only, we
obtain the bound given in (\ref{eq:bound on most general f}), 
\begin{equation}
f^{L}(x,y,w)=\frac{2\left(\mu\nu-\lambda^{2}\right)}{\nu}x=2e_{n}\hbar\sqrt{\mu\nu-\lambda^{2}}\geq\hbar\sqrt{\mu\nu-\lambda^{2}}\,.
\end{equation}
Up to phase-space translations $\hat{T}_{\alpha}$, a single squeezed
state saturates the bound, namely 
\begin{equation}
{\cal M}(f^{L})=\left\{ \ket{\mu,\nu,\lambda}=\hat{G}_{\frac{\lambda}{\nu}}\hat{S}_{\frac{1}{2}\ln\left(\frac{\nu}{\sqrt{\mu\nu-\lambda^{2}}}\right)}\ket 0\right\} .
\end{equation}

When expressing the correlation term $C_{pq}$ in terms of the variance
$\Delta^{2}r$ according to Eq.~(\ref{eq: variance of r}), we obtain,
for $\mu=\nu=2\lambda=1$ in (\ref{eq:bound on most general f}),
the \emph{triple sum} uncertainty relation 
\begin{equation}
\Delta^{\!2}p+\Delta^{\!2}q+\Delta^{\!2}r\geq\sqrt{3}\hbar\,,\label{eq: triple sum inequality}
\end{equation}
derived in \cite{kechrimparis+14}, and the minimum is achieved for
the state $\ket{1,1,1/2}\equiv\ket{\Xi_{0}}$ which also minimizes
the triple product uncertainty (cf. Eq.~(\ref{eq: Xi_0}) and Fig.~\ref{fig: triple-uncertainty-visual}).

\emph{Sums of powers} of position and momentum variances are bounded
from below according to the inequality 
\begin{equation}
\mu\left(\Delta^{2}p\right)^{m}+\nu\left(\Delta^{2}q\right)^{m^{\prime}}\geq\left(\frac{\hbar}{2}\right)^{\frac{2mm^\prime}{m+m^{\prime}}}\left(\mu\left(\frac{\nu}{\mu}\frac{m^{\prime}}{m}\right)^{\frac{m}{m+m^{\prime}}}+\nu\left(\frac{\mu}{\nu}\frac{m}{m^\prime}\right)^{\frac{m^\prime}{m+m^{\prime}}}\right)\,,\qquad m,m^{\prime}\in\mathbb{N}\,,\label{eq: sums of powers}
\end{equation}
reducing to the \emph{pair sum} uncertainty relation (\ref{eq: sum inequality})
in the simplest case ($\mu=\nu=m=m^{\prime}=1$). 

Next, we study a \emph{generalized RS-uncertainty functional }(\ref{eq: RS function}),
\begin{equation}
f_{m,m^{\prime}}^{RS}(x,y,w)=\left(xy\right)^{m}-\mu w^{m^{\prime}}\,,\qquad\mu>0\,,\quad m,m^{\prime}\in\mathbb{N}\,.\label{eq: fgRS}
\end{equation}
 For arbitrary integers $m$ and $m^\prime$, the consistency conditions
cannot be solved in closed form. Setting $m^{\prime}=2m$ and assuming
that both $m>1$ and $\mu>1$ hold, we obtain the explicit bound 
\begin{equation}
\left(\Delta^{2}p\cdot\Delta^{2}q\right)^{m}-\mu\left(C_{qp}\right)^{2m}\geq\left(\frac{\hbar}{2}\right)^{2m}\frac{\mu}{\left(\mu^{\frac{1}{m-1}}-1\right)^{m}}\,.
\end{equation}
An interesting special case of $f_{m,2m}^{RS}$ occurs for $m=\nicefrac{1}{2}$
and $0<\mu<1$, 
\begin{equation}
\Delta p\Delta q-\mu\left|C_{pq}\right|\geq\frac{\hbar}{2}\sqrt{1-\mu^{2}}\,,\label{eq: RS modified}
\end{equation}
which can be treated in spite of the presence of the non-differentiable
term. The extremal states depend on one free parameter, 

\begin{equation}
{\cal E}_{\alpha}(f_{1,\nicefrac{1}{2}}^{RS})=\left\{ \ket{\alpha,n}=\hat{T}_{\alpha}\hat{G}_{\pm\frac{\mu e_{n}\hbar}{2x\sqrt{1-\mu^{2}}}}\hat{S}_{\frac{1}{2}\ln\left(\frac{x}{e_{n}\hbar}\right)}\ket n\right\} \,,\qquad x>0\,;\label{eq: generalized RS}
\end{equation}
in the absence of the covariance term, $\mu=0$, they reduce to the
squeezed number states with a real parameter known to extremize Heisenberg's
inequality.

Next, we present an example of an uncertainty relation which seems
to be entirely out of reach of traditional derivations. Defining the
functional 
\begin{equation}
f^{e}(x,y)=x+\mu e^{y/\nu}\,,\quad\mu,\nu>0\,,
\end{equation}
we obtain the inequality 
\begin{equation}
\Delta^{2}p+\mu e^{\Delta^{2}q/\nu}\geq\left(1+2W(\hbar/4\sqrt{\mu\nu})\right)e^{2W(\hbar/4\sqrt{\mu\nu}))}\,,\label{eq: exponential inequality}
\end{equation}
using the fact that Lambert's $W$-function $W(s)$, defined as the
inverse of $s(W)=W\exp W$, is a strictly increasing function. In
the limit of $\mu\to\infty$ and assuming $\mu=\nu$, the left-hand-side
of \eqref{eq: exponential inequality} turns into $(\mu+\Delta^{\!2}p+\Delta^{\!2}q+{\cal O}(1/\mu))$
while the expansion of its right-hand-side produces the correct bound
$(\mu+\hbar+{\cal O}(1/\mu))$, since $W(s)=s+{\cal O}(s^{2})$.

The position and momentum variances at the extremum with label $n\in\mathbb{N}_{0}$
are given by 
\begin{align}
x & =2\mu W\left(\frac{e_{n}\hbar}{2\sqrt{\mu\nu}}\right)e^{W\left(\frac{e_{n}\hbar}{2\sqrt{\mu\nu}}\right)}
\end{align}
and 

\begin{equation}
y=2\nu W\left(\frac{e_{n}\hbar}{2\sqrt{\mu\nu}}\right)\,,
\end{equation}
respectively. Using Eqs.~\eqref{eq: definitions of b and gamma}
with $\det\mathbf{F}=(\mu/\nu)e^{\nicefrac{y}{\nu}}$, one finds 
\begin{equation}
b=0\,,\qquad\gamma=\frac{1}{4}\ln\left(\frac{\mu}{\nu}\right)+\frac{1}{2}W\left(\frac{e_{n}\hbar}{2\sqrt{\mu\nu}}\right)\,,
\end{equation}
which means that only a single state (and its rigid displacements)
will saturate the inequality \eqref{eq: exponential inequality}.
If $\mu=\nu$, we recover $x=y=\hbar^{2}/4$ as well as $b=\gamma=0$,
i.e. the ground state of a unit oscillator since $W(0)=0$.

Let us point out that some general statements can be made about
functionals of the form $f=f(xy,w)$ and $f=f(\mu x^{m}+\nu y^{m^{\prime}},w)$,
i.e. generalizations of the expressions in \eqref{eq: sums of powers}
and \eqref{eq: fgRS}, respectively. By examining the consistency
conditions one can show that the extrema of the first expression come
as a one-parameter set, while they are isolated or a one-parameter
family in the second case. However, without knowing the explicit form
of the functions no further conclusions can be drawn. 

\emph{Rational function}s of the variances such us
\begin{equation}
f_{\mu,\nu}^{r}(x,y)=\frac{x^{m}y^{m}}{\mu x^{m^{\prime}}+\nu y^{m^{\prime}}}\,,\quad\mu,\nu>0\,,\label{eq: rational functions}
\end{equation}
lead to an interesting class of uncertainty relations. If $m=m^\prime=\nicefrac{1}{2}$, for example, the extremal states of the functional are given by the set
\begin{equation}
{\cal M}_{\alpha}(f_{\nicefrac{1}{2},\nicefrac{1}{2}}^{r})=\left\{ \ket{\alpha,n}=\hat{T}_{\alpha}\hat{S}_{\frac{1}{2}\ln\left(\frac{\nu}{\mu}\right)}\ket n\right\} .\label{eq: rational family}
\end{equation}
However, none of these states minimizes the uncertainty functional. In any family of squeezed states with real squeezing parameter, the \emph{product} of the position and momentum variances is constant while their \emph{sum} increases without a bound. Consequently, the functional can be made arbitrarily small without ever reaching the value zero.

\subsection{Uncertainty functionals with permutation symmetries\label{subsec: j=00005Bpsi=00005D with discrete symmetries}}

The triple product uncertainty relation and the one derived by Heisenberg
possess discrete symmetries. Here we investigate more general uncertainty
functionals which are invariant under the exchange of three and two
variances. 

\subsubsection*{$S_{3}$-invariant functionals }

Consider a function of three variables which is invariant under the
exchange of any pair, 

\begin{equation}
f^{(3)}(x,y,z)=f^{(3)}(y,x,z)=f^{(3)}(x,z,y)\,.\label{eq: 3fold symetric functional}
\end{equation}
We now derive the lower bound of a large class of $S_{3}$-invariant
uncertainty functionals $J[\psi]$ and show that their minima coincide
with the state $\ket{\Xi_{0}}$ minimizing the triple product inequality.
The variables $x,y$ and $z$ will denote the variances of the operators
$\hat{p},\hat{q}$ and $\hat{r}$, respectively. 

More specifically, we study the minima of sums of completely homogeneous
polynomials of degree $n$, with arbitrary \emph{non-negative} coefficients,
\begin{equation}
f_{N}^{(3)}(x,y,z)=\sum_{n=1}^{N}\,\sum_{j+k+\ell=n}a_{jk\ell}x^{j}y^{k}z^{\ell}\,,\quad a_{jk\ell}\geq0\,,\label{eq: def f(3)N}
\end{equation}
dropping the unimportant constant term $a_{000}$. The determinant
of the associated $\mathbf{F}$-matrix, 

\begin{equation}
\det\,\mathbf{F}\equiv f_{x}f_{y}+f_{y}f_{z}+f_{z}f_{x}\,,
\end{equation}
is positive definite since $x,y,z>0$ and the partial derivatives
are just positive polynomials.

The symmetry under $S_{3}$-permutations \eqref{eq: 3fold symetric functional}
implies that the coefficients must satisfy the conditions 
\begin{align}
a_{jk\ell}=a_{kj\ell}=a_{j\ell k}\,,\quad0\leq j,k,\ell\leq n\,,\label{SymmetryConditions}
\end{align}
so that the first terms of the polynomials are given by
\begin{eqnarray}
f_{N}^{(3)}(x,y,z) & = & a_{100}(x+y+z)\nonumber \\
 &  & +a_{200}\left(x^{2}+y^{2}+z^{2}\right)+a_{110}\left(xy+yz+zx\right)\nonumber \\
 &  & +a_{300}\left(x^{3}+y^{3}+z^{3}\right)+a_{210}\left(x^{2}(y+z)+y^{2}(z+x)+z^{2}(x+y)\right)\nonumber \\
 &  & +a_{111}xyz+\dots\label{eq: explicit f(3)N terms}
\end{eqnarray}
If the only nonzero coefficients are $a_{100}=1$ or $a_{111}=1$,
we recover the functionals associated with the triple sum \eqref{eq: triple sum inequality}
or the triple product inequality \eqref{eq: triple product uncertainty},
respectively. In general, a completely homogeneous $S_{3}$-symmetric
polynomial in three variables of degree $n\geq1$ consists of $\kappa_{n}=\left\lfloor \frac{\left(n+3\right)^{2}+6}{12}\right\rfloor $
terms where the floor function $\left\lfloor s\right\rfloor $ denotes
the integer part of the number $s$: each term arises from one way
to partition $j+k+\ell=n$ objects into three sets with $j,k$ and
$\ell$ elements, respectively \cite{sloane+95}. Thus, a symmetric
polynomial of degree up to $N$ depends on $\left(\sum_{n=1}^{N}\kappa_{n}\right)$
independent coefficients if one ignores the constant term. 

The main result of this section follows from rewriting the consistency
conditions \eqref{eq: alternative consistency I} and \eqref{eq: alternative consistency RS part}
in terms of the variables $x,y$ and $z$, 
\begin{align}
xf_{x}-yf_{y}+\left(x-y\right)f_{z} & =0\,,\label{eq: Extremal conditions2 a}\\
zf_{z}-xf_{x}+\left(z-x\right)f_{y} & =0\,,\label{eq: ExtremalConditions2 b}\\
2\left(xy+yz+zx\right)-x^{2}-y^{2}-z^{2} & =\left(2n+1\right)^{2}\hbar^{2}\,,\label{eq: ExtremalConditions2 c}
\end{align}
where we have used the identity $z=x+y+2w$ given in \eqref{eq: variance of r}.
The conditions \eqref{eq: Extremal conditions2 a} and \eqref{eq: ExtremalConditions2 b}
imply that the extrema of any symmetric polynomial $f_{N}^{(3)}(x,y,z)$
occur whenever the three variances take the same value, 
\begin{equation}
x=y=z\,.\label{eq: s3-implied equality of variances}
\end{equation}
To show that $x=y$ holds we pick any nonzero term $a_{jk\ell}x^{j}y^{k}z^{\ell}$
in the expansion \eqref{eq: def f(3)N} and assume that the powers
of $x$ and $y$ are different, i.e. $j\neq k$; the case $j=k$ will
be considered later. Due to the symmetry under the exchange $x\leftrightarrow y$,
the sum also must contain the term $a_{kj\ell}x^{k}y^{j}z^{\ell}$,
with $a_{kj\ell}\equiv a_{jk\ell}$. Defining $t(x,y,z)=a_{jk\ell}\left(x^{j}y^{k}+x^{k}y^{j}\right)z^{\ell}$,
the first two terms of \eqref{eq: Extremal conditions2 a} take the
form 
\begin{align}
xt_{x}-yt_{y} & =(j-k)a_{jkl}\left(x^{j}y^{k}-x^{k}y^{j}\right)z^{\ell}\,.\label{eq: first part of CCond1-1}
\end{align}
Assuming that $j=k+\delta,$ with $\delta>0$, we find
\begin{align}
xt_{x}-yt_{y} & =a_{k+\delta\,kl}\delta\left(x^{\mu}-y^{\mu}\right)x^{k}y^{k}z^{\ell}\\
 & =(x-y)\delta a_{k+\delta\,k\ell}\left(x^{\delta-1}+x^{\delta-2}y+\ldots+xy^{\delta-2}+y^{\delta-1}\right)z^{\ell}\\
 & \equiv(x-y)g_{+}(x,y,z)\,,
\end{align}
where $g_{+}(x,y,z)>0$. Using this expression in \eqref{eq: Extremal conditions2 a},
the consistency condition takes the form
\begin{equation}
(x-y)\left(g_{+}(x,y,z)+t_{z}\right)=0\,,
\end{equation}
with another positive function $t_{z}(x,y,z)$. If $\delta<0$ we
write $k=j-\delta\equiv j+\left|\delta\right|$ and eliminate $k$
instead of $j$ from \eqref{eq: first part of CCond1-1}, only to
find that its left-hand-side again turns into $(x-y)$ multiplied
with a positive function. If the powers of $x$ and $y$ of the term
$a_{kj\ell}x^{k}y^{j}z^{\ell}$ are equal, $j=k$, one immediately
finds that $(x\partial_{x}-y\partial_{y})a_{jkl}x^{j}y^{j}z^{\ell}=0$,
also reducing Eq.~\eqref{eq: Extremal conditions2 a} to $(x-y)\partial_{z}a_{jkl}x^{j}y^{j}z^{\ell}=0$. 

The argument just given covers \emph{all} terms in the sum \eqref{eq: def f(3)N},
and the positivity of the coefficients $a_{jk\ell}$ implies that
the first consistency condition can only be satisfied for $x=y$.
Using the symmetry of $f_{N}^{(3)}(x,y,z)$ under the exchange $y\leftrightarrow z$,
an identical argument leads to the identity $y=z$.

Using \eqref{eq: s3-implied equality of variances} to evaluate the
left-hand-side of Eq. \eqref{eq: ExtremalConditions2 c} results in
\begin{equation}
x=y=z=\tau e_{n}\hbar\,,\qquad n\in\mathbb{N}_{0}\,,\label{Extrema}
\end{equation}
where $\tau=\sqrt{4/3}$, so that we obtain an uncertainty relation
for any $S_{3}$-invariant function
\begin{equation}
f_{N}^{(3)}(x,y,z)\geq\left.f_{N}^{(3)}(x,y,z)\right|_{x=y=z=\hbar/\sqrt{3}}\,.\label{eq: s3-minimum}
\end{equation}
This result correctly reproduces the special cases of Eqs.~\eqref{eq: triple product uncertainty}
and \eqref{eq: triple sum inequality}, and there is only one state
which saturates the inequality, namely $\ket{\Xi_{0}}$ given in Eq.~\eqref{eq: Xi_0}.
Letting $N\rightarrow\infty$ in Eq.~\eqref{eq: def f(3)N}, we conclude
that the main result of this section, Eq.~\eqref{eq: s3-minimum},
also applies to any $S_{3}$-symmetric function $f_{\infty}^{(3)}(x,y,z)$
with a Taylor expansion with positive coefficients and infinite radius
of convergence, as long as its first partial derivatives exist.

\subsubsection*{$S_{2}$-invariant functionals}

Assume now that, in analogy to Eq.~\eqref{eq: 3fold symetric functional},
we have a functional depending on just two variances in a symmetric
way, 

\begin{equation}
f^{(2)}(x,y)=f^{(2)}(y,x)\,.\label{eq: 2fold symetric functional}
\end{equation}
An argument similar to the one given for the function $f^{(3)}(x,y,z)$
results in the uncertainty relation
\begin{equation}
f_{N}^{(2)}(x,y)\geq\left.f_{N}^{(2)}(x,y)\right|_{x=y=\hbar/2}\,,\label{eq: s2-minimum}
\end{equation}
which covers the cases of Heisenberg's relation \eqref{eq: Heisenbergs inequality}
and the pair sum inequality \eqref{eq: sum inequality}. Thus, the
actual form of the function at hand determines whether the set of
minima ${\cal M}(f_{N}^{(2)})$ will depend on a continuous parameter
or not. If the functional is invariant under scaling transformation
$x\to\lambda x$, $y\to y/\lambda$, in addition to the permutation
symmetry, there is a one-parameter family of solutions and the right-hand-side
of Eq.~\eqref{eq: s2-minimum} achieves its minimum on the set of
points with $xy=(\hbar/2)^{2}$, not just those with $x=y=\hbar/2$.

To derive \eqref{eq: s2-minimum}, we suppose that the function $f_{N}^{(2)}(x,y)$
has an expansion in analogy to $f_{N}^{(3)}(x,y,z)$ in Eq.~\eqref{eq: def f(3)N}
but without the variable $z$. Adapting the reasoning applied to $f_{N}^{(3)}(x,y,z)$,
the consistency equations \eqref{eq: alternative consistency I} are
found to imply $x=y$ and $w=0$. Using this result in \eqref{eq: alternative consistency RS part},
the bound $x\geq\hbar^{2}/4$ follows immediately, so that the inequality
\eqref{eq: s2-minimum} must hold for $S_{2}$-invariant functionals.

\section{Summary and discussion\label{sec:Summary-and-concluding}}

This paper responds to the fact that, without exception, the states
minimizing the known preparational uncertainty relations for a quantum
particle in one dimension are given by (sets of) squeezed states.
To explain this fact we systematically study lower bounds for smooth
functions $f(\Delta^{2}p,\Delta^{2}q,C_{pq})$ of second moments.
The resulting theory explains the universal role of squeezed states
for preparational uncertainty relations depending on second moments
only, and it completely charts the landscape of inequalities of this
type.

The chain of inclusions 
\begin{equation}
{\cal E}\supseteq{\cal E}(f)\supseteq M(f)\,\label{eq: inclusions without alpha}
\end{equation}
concisely summarizes the general structure of our findings. First,
we have shown that only squeezed number states of a quantum mechanical
harmonic oscillator with unit frequency and mass occur as extrema
of an uncertainty functional $J[\psi]$ depending on second moments.
We denote this \emph{universal} set of states by ${\cal E}$. Second,
the extrema of a \emph{specific} functional $J_{f}[\psi]$, associated
with a function $f(\Delta^{2}p,\Delta^{2}q,C_{pq})$, form the subset
${\cal E}(f)$ of the universal set ${\cal E}$. Third, the functional
will assume its minimum for one or more of the extrema ${\cal E}(f)$,
a subset which we denote by ${\cal M}(f)$. The set of minima may
be empty, ${\cal M}(f)=\textrm{Ø}$. If it is not empty, a lower bound
on the functional $J_{f}[\psi]$ has been found, and it represents
a preparational uncertainty relation in terms of the second moments. 

Strictly speaking, we obtained the relations ${\cal E}_{\alpha}\supseteq{\cal E}_{\alpha}(f)\supseteq{\cal M}_{\alpha}(f)$
instead of Eq.~\eqref{eq: inclusions without alpha} as it is possible
to move quantum states in phase space without affecting the values
of the second moments. The four-parameter set ${\cal E}_{\alpha}\equiv\hat{T}_{\alpha}{\cal E}$,
for example, consists of the squeezed states ${\cal E}$ plus those
obtained from them by means of the translation operator $\hat{T}_{\alpha}$
defined in Eq. \eqref{TranslOp}. Thus, each state saturating a specific
inequality with vanishing expectation values gives rise to a two-parameter
family of minima.

Our results have a useful geometric representation in the real three-dimensional
space of moments. The uncertainty region, consisting of all triples
of moments which can arise from (pure or mixed) states of a quantum
particle, turns out to be a convex set bounded by a one-sheeted hyperboloid.
The boundary is invariant under the elliptic rotations, hyperbolic
boosts and parabolic transformations which generate the group $\mbox{SO}(1,2)$.
This observation squares with the importance of the group $\mbox{SO}(1,2)$
in quantum optics where coherent and squeezed states are ubiquitous. 

The invariance of the bounding hyperboloid can, in turn, also be understood
as an invariance of the functional $J[\psi]$ defining the surface.
Each point on this hyperboloid is associated with a unique Gaussian
state saturating the Robertson-Schrödinger inequality. Changing from
an active view of transformations (i.e. mapping one state with minimal
uncertainty to another one) to a passive view, we see that the RS-functional
(or a suitable smooth function thereof) is invariant under the elements
of the group $\mbox{Sp}(2,\mathbb{R})$ applied to the canonical pair
$(\hat{p},\hat{q})$. The allowed symplectic transformations include
rotations, scalings and linear gauge transformations, or shears. 

Repeatedly, we have encountered subsets of the maximal possible symmetry
group $\mbox{Sp}(2,\mathbb{R})$ (cf. \cite{trifonov01}). The Heisenberg
functional $f^{H}(x,y,z)$ in Eq.\@ \eqref{eq: HB functional}, for example,
is invariant under a scaling transformation $\hat{p}\to\lambda\hat{p}$,
$\hat{q}\to\hat{q}/\lambda$, with $\lambda>0$, resulting in a one-parameter
set of states with minimal uncertainty depicted in Fig. \ref{fig:HB extrema}.
Similarly, uncertainty functionals invariant under permutations of
order two or three are minimized by states with corresponding symmetry
properties. Examples are the triple product uncertainty relation \eqref{eq: triple product uncertainty}
and, more generally, the functionals with discrete symmetries discussed
in Sec. \ref{subsec: j=00005Bpsi=00005D with discrete symmetries}.

\begin{figure}
\begin{centering}
\def\svgwidth{0.7\textwidth}
\begin{center}
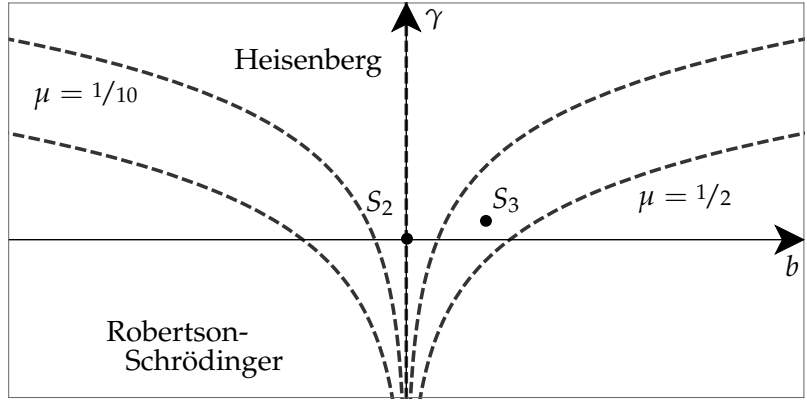
\end{center}
\par\end{centering}
\centering{}\caption{\label{fig: all uncertainty relations} States on the boundary of
the uncertainty region minimizing known and new uncertainty relations
parameterized by the real numbers $(b,\gamma)$, with $\hbar=1$;
each point of the plane corresponds to a squeezed state saturating
the RS-inequality \eqref{eq: RS inequality}; points on the vertical
dashed line represent minima of Heisenberg's uncertainty relation
\eqref{eq: Heisenbergs inequality}; the two curved dashed lines indicate
the minima of the modified RS-inequality \eqref{eq: RS modified}
with $m=1/2$ and values $\mu=1/2$ (bottom) and $\mu=1/10$ (top);
the full dots correspond to minima of $S_{2}$-invariant functionals
\eqref{eq: s2-minimum} such as the pair sum \eqref{eq: sum inequality}
and $S_{3}$-invariant functionals \eqref{eq: s3-minimum} such as
the triple product \eqref{eq: triple product uncertainty}.}
\end{figure}

We also derived new and explicit uncertainty relations. Fig.~\ref{fig: all uncertainty relations}
uses the $(b,\gamma)$-plane to illustrate the sets of states which
minimize (some of) the uncertainty relations discussed in this paper.
The sets of minima ${\cal M}$ may depend on \emph{two} parameters
(all squeezed states minimizing the RS-inequality), on \emph{one}
parameter (such as the real squeezed states saturating Heisenberg's
uncertainty relation) or consist of a \emph{single point} only (associated
with $S_{3}$-invariant inequalities such as the triple product inequality,
for example). We have not been able to backward-engineer functionals
which would be minimized by prescribed subsets of the plane such as
a circle or a disk. The minimizing states we found were all pure,
located on the \emph{boundary} of the uncertainty region. In principle,
functionals could also take their minima \emph{inside} this region
although we have found only trivial examples with this property \cite{kechrimparis+16}. 

We currently investigate three conceptually interesting generalizations
of our approach. First, there is no fundamental reason to restrict
oneself to uncertainty functionals depending only on second moments
in position and momentum \cite{busch+14}. On the contrary, higher
order expectation values would enable us to move away from Gaussian
quantum mechanics which is largely reproducible in terms of a classical
model ``with an epistemic restriction'' of the allowed probability
distributions \cite{bartlett+12}. Including fourth-order terms $\bra{\psi}\hat{q}^{4}\ket{\psi}$,
for example, will result in an eigenvalue equation \eqref{EigEq}
which is not related to a unit oscillator in a simple way. It is known
that fourth-order moments for single-particle expectations can give
rise to inequalities which cannot be reproduced by models based on
classical probabilities \cite{bednorz+11,kot+12}. Thus, it might
become possible to study truly non-classical behaviour in a systematic
manner using suitable uncertainty functionals.

Secondly, our approach can be generalized to the case of two or more
continuous variables. We expect that a systematic study of uncertainty
functionals becomes possible, leading to criteria which would detect
pure entangled states. For example, a generalization of the triple
uncertainty relation \eqref{eq: triple product uncertainty} to bipartite
systems has been shown to detect entangled states in a quantum optical
setting \cite{paul+16}. Other scenarios are known which rely on intuitive
choices of suitable bi-linear observables \cite{duan+00,serafini06}. 

Finally, the comprehensive study \cite{dammeier+15} of uncertainty
relations for a single spin $s$ has been limited to observables transforming
covariantly under the group $\mbox{SU}(2)$. The method proposed here
is easily adapted to investigate functionals depending on arbitrary
functions of moments.

\subsection*{Acknowledgments}

The authors would like to thank Reinhard F. Werner for comments on
an early draft of this paper and, more generally, for discussions
of uncertainty relations. Tom Bullock kindly commented on a late version
of our manuscript. S.K. has been supported via the act ``Scholarship
Programme of S.S.F.\@  by the procedure of individual assessment, of 2011\textendash 12''
by resources of the Operational Programme for Education and Lifelong
Learning of the ESF and of the NSF, 2007\textendash 2013.

\appendix

\section{A Baker-Campbell-Hausdorff identity\label{sec: An-operator-identity}}

The relation $\hat{G}_{b}\hat{S}_{\gamma}=\hat{S}(\xi)\hat{R}(\chi)$
in Eq.~\eqref{eq: BCH identity: GS =00003D SR} can be shown by requiring
that both products map the annihilation operator $\hat{a}=\left(\hat{q}+i\hat{p}\right)/\sqrt{2\hbar}$
to the same operator. We obtain
\begin{equation}
\hat{G}_{b}\hat{S}_{\gamma}\hat{a}\hat{S}_{\gamma}^{\dagger}\hat{G}_{b}^{\dagger}=\hat{a}\left(\cosh\gamma-i\frac{b}{2}e^{\gamma}\right)+\hat{a}^{\dagger}\left(\sinh\gamma+i\frac{b}{2}e^{\gamma}\right)
\end{equation}
and 
\begin{equation}
\hat{S}(\xi)\hat{R}(\chi)\hat{a}\hat{R}^{\dagger}(\chi)\hat{S}^{\dagger}(\xi)=\hat{a}e^{-i\chi}\cosh r-\hat{a}^{\dagger}e^{-i\chi}e^{i\theta}\sinh r\,,
\end{equation}
respectively. Equating the coefficients of the operators $\hat{a}$
and $\hat{a}^{\dagger}$ leads to two equations 
\begin{align}
\cosh\gamma-i\frac{b}{2}e^{\gamma} & =e^{-i\chi}\cosh r\,,\\
\sinh\gamma+i\frac{b}{2}e^{\gamma} & =-e^{i(\theta-\chi)}\sinh r\,,\label{soe}
\end{align}
which we need to solve for the variables $\xi\equiv re^{i\theta}$
and $\chi$. Separating the real and imaginary parts of the first
equation, one finds that 
\begin{equation}
\chi=\arctan\left(\frac{b}{1+e^{-2\gamma}}\right)\in(-\frac{\pi}{2}\,,\frac{\pi}{2})\,.\label{eq: solution for chi}
\end{equation}
In a similar way, the second equation allows one to solve for the
function $\tan(\theta-\chi)$ which, upon using \eqref{eq: solution for chi},
leads to

\begin{equation}
\theta=\arctan\left(\frac{b}{1-e^{-2\gamma}}\right)+\arctan\left(\frac{b}{1+e^{-2\gamma}}\right)\in(-\pi\,,\pi)\,.\label{theta}
\end{equation}
The case of $\gamma=0$ needs to be treated separately leading to
the relation 
\begin{equation}
\theta=\pm\frac{\pi}{2}+\arctan\left(\frac{b}{2}\right)\,\in(-\pi,\pi).
\end{equation}
Finally, the condition $\cosh r\,\cos\chi=\cosh\gamma$ results in
the expression 
\begin{equation}
r=\text{arcosh}\left(\cosh^{2}\gamma+\frac{b^{2}}{4}e^{2\gamma}\right)^{1/2}\,\in[0,\infty)\,,\label{r}
\end{equation}
which establishes the desired identity \eqref{eq: BCH identity: GS =00003D SR}. 

The number states $\ket n,n\in\mathbb{N}$, are eigenstates of phase-space
rotations $\hat{R}(\chi)$. Therefore, the product $\hat{G}_{b}\hat{S}_{\gamma}$
acts on those states according to 
\begin{equation}
\hat{G}_{b}\hat{S}_{\gamma}\ket n\cong\hat{S}(\xi)\ket n\,,
\end{equation}
where an irrelevant phase has been suppressed. Thus, the operator
$\hat{S}(\xi)$ generates all squeezed states from $\ket 0$ when
the parameter $\xi$ runs through the points of the complex plane.

\end{document}